\documentclass[aps,prl,reprint,superscriptaddress]{revtex4-1}

\usepackage{amsmath,amssymb}
\usepackage{graphicx}
\usepackage{bm}
\usepackage[latin1]{inputenc}

\usepackage[caption=false]{subfig}

\RequirePackage{hyperref}

\usepackage{color}
\usepackage{soul}

\begin{document}

\title{Anti-alignment driven dynamics in the excited states of molecules under strong fields}

\author{Sebasti\'an Carrasco}
\email[]{sebastian.carrasco@ug.uchile.cl}
\affiliation{Departamento de F\'isica, Facultad de Ciencias,
  Universidad de Chile, Casilla 653, Santiago, Chile 7800024} 
\affiliation{Centro para la Nanociencia y la Nanotecnolg\'ia, CEDENNA,
  Chile} 
  
\author{Jos\'e Rogan}
\affiliation{Departamento de F\'isica, Facultad de Ciencias,
  Universidad de Chile, Casilla 653, Santiago, Chile 7800024} 
\affiliation{Centro para la Nanociencia y la Nanotecnolg\'ia, CEDENNA,
  Chile}

\author{Juan Alejandro Valdivia}
\affiliation{Departamento de F\'isica, Facultad de Ciencias,
  Universidad de Chile, Casilla 653, Santiago, Chile 7800024} 
\affiliation{Centro para la Nanociencia y la Nanotecnolg\'ia, CEDENNA,
  Chile} 
  
\author{Ignacio Sola}
\affiliation{Departamento de Qu\'{\i}mica F\'{\i}sica, Universidad Complutense, 28040 Madrid, Spain.}

\date{\today}

\begin{abstract}
  \noindent We develop two novel models of the H$_2^+$ molecule and its isotopes from which we assess quantum-mechanically and semi-classically whether the molecule anti-aligns with the field in the first excited electronic state. The results from both models allow us to predict anti-alignment dynamics even for the HD$^+$ isotope, which possesses a permanent dipole moment. The molecule dissociates at angles perpendicular to the field polarization in both the excited and the ground electronic state, as the population is exchanged through a conical intersection. The quantum mechanical dispersion of the initial state is sufficient to cause full dissociation. We conclude that the stabilization of these molecules in the excited state through bond-hardening under a strong field is highly unlikely.
\end{abstract}


\maketitle

\section{Introduction}
Quantum control in molecules is particularly challenging for the interplay of different degrees of freedom in the coherent response of the system: electronic, vibrational, and rotational\cite{Rice2000,Brumer2003,BrifACP12,WorthARC13,SolaAAMO18}. This is especially the case when the external action has to be maintained during longer times and involves excited states of the molecule. From the theoretical point of view, it is almost impossible to study the system fully quantum-mechanically. When this is done, it requires separating the different motions. However, under strong fields, when ionization and dissociation can take place and conical intersections are pervasive, it is important to treat the system as much as possible in a fully consistent way. In this scenario, the study of the dynamics of H$_2^+$ or its isotopes is a theoretical lab where different approaches and control ideas can be tested. In addition, the lightness of the protons makes the interplay of the different degrees of freedom more important to analyze.

To date, most studies have focused on some degrees of freedom at a time. Seminal work has shown both theoretically and experimentally how molecules in their ground states align or orient to an external field, depending on whether the molecule possesses a permanent dipole or the effect is induced by the polarizability\cite{StapelfeldtRMP03,SeidemanAAMO05}. 
On the other hand, under strong optical or near-infrared pulses, it was shown that the molecular potentials change leading to bond softening in the ground electronic state\cite{BucksbaumPRL90,ZavriyevPRL93,Giusti-SuzorJPB95} and bond hardening in excited dissociative states\cite{Giusti-SuzorPRL92,YaoCPL92,ZavriyevPRL93,AubanelPRA93,ChangCPC13}. These effects can be naturally explained in the adiabatic representation that diagonalizes the Hamiltonian including the coupling with the external field, using the Rotating-wave approximation (RWA) or a Floquet representation, by which the energy of the molecular potentials is first shifted by the photon energies. 
Bond softening and bond hardening are then manifestations of the mixing of the bonding and anti-bonding orbitals that characterize the ground and excited molecular orbitals under the coupling of the field. In these studies, the assumption that the molecule is aligned, of at least that its axis remains mostly fixed with respect to the polarization field, has been implicitly used in the proposal of many quantum control scenarios using strong non-resonant fields\cite{BalintKurtiACP08,ChangIJQC16}, which are based on the dynamic Stark effect\cite{SussmanSci06,TownsendJPCA11,KimJPCA12,CorralesNat14,SolaPCCP15}.

The general interest has spiked as experimental advances have  allowed to study the structural features of the molecule under these fields, the so-called light-induced potentials (LIPs) \cite{YuanJCP78, bandrauk1981photodissociation, GarrawayPRL98,SolaPRL00,EmilNJP09}, reaching the point of analyzing interesting topological effects, such as light-induced conical intersections (LICIs) \cite{moiseyev2008laser, halasz2012light, demekhin2013light, halasz2015direct, CsehiJPCL17}. But because all these structures lie far from equilibrium, it becomes increasingly important to study the dynamics of the molecule considering all its degrees of freedom. It is in this context that we have recently developed simplified models to solve the electronic and nuclear (vibration and rotation) motion of H$_2^+$ in strong fields\cite{chang2019grid}. These models used soft-core Coulomb potentials\cite{javanainen1988numerical, su1991model, kulander1996model} in a plane (instead of a line) where the nuclear motion was treated using the Ehrenfest ansatz. 

In this work, we work with constant fields, rather than laser pulses, in order to simplify the analysis and better isolate the different effects that arise from the field, without invoking the RWA or using a Floquet representation. Then one can obtain the true adiabatic states of the Hamiltonian, which we call field-induced potentials (FIPs), or molecular states dressed by virtual photons, instead of LIPs. One can expect that many of the effects provoked by strong constant fields will be applicable for non-resonant lasers, as the coupling in both cases is induced by the polarizability. However, since with constant fields the potentials are not shifted by the energy of the photons, there cannot be proper light-induced conical intersections, but avoided crossings. And, as the energy differences between electronic states that dissociate in the same chemical channels converge asymptotically, it is interesting to study possible electronic transitions induced by the nuclear motion at large internuclear distances, that play the role of non-adiabatic crossings between the FIPs.

Under a strong constant field, we found that the dynamics in the excited state of H$_2^+$ leads to anti-alignment instead of the typical alignment that is mostly assumed in simplified models\cite{chang2019control}. While the excited state has a deep well in the presence of the strong field when the molecular axis is aligned with the field, and thus one observes vibrational motion, the molecules that are not initially aligned experience a torque that rotates the molecule perpendicular to the field, where the coupling is zero, and hence the molecule dissociates in the excited state. The main consequence of this finding is that it becomes impossible to use the excited states dressed by the field to control the molecular bond\cite{ChangJCP03,ChangPRA03,ChangJCP13}.  On the other hand, using laser pulses, it was theoretically predicted that there is a LICI between the ground and first excited electronic states at large internuclear distances when the axis is perpendicular to the field\cite{SindelkaJPB11,Halasz2013Effect, halasz2013nuclear, kubel2020probing,HalaszJPCA14}. Therefore, anti-alignment can be used as a resource to explore the dynamics in the proximity of LICIs. However, alignment in the ground state leads to larger bond softening and possibly dissociation as well. One can then also determine whether the dissociation occurs mainly in the excited state at $\pm 90^\circ$, or in the ground state, at $\theta = 0$, due to bond softening, as reported in several experiments\cite{ZavriyevPRA90,AubanelPRA93,NumicoPRA99,NatanPRL16}. One can also expect an even faster nuclear wave packet dephasing (or rotationally-induced decoherence) as the molecular axis disperses following opposite directions in the ground and excited state.

All these questions can not be solved until a full quantum treatment of all the molecular degrees of freedom is developed, preferentially using the true Coulomb potential. Only then one can fully assess whether the predicted anti-alignment dynamics is an artifact of the approximations (dimensional reduction, soft-core potentials) or not. 

In this work, we develop two novel models. In the first one, we combine the Ehrenfest approach for nuclear motion with the well-established prolate spheroidal treatment for an electron in a two-center molecule~\cite{baik1996multiphoton,kamta2004high,kamta2005three}. In the last one, we solve the three coordinates of the electronic motion using the same treatment as before, combining it with the second-order split-operator method to solve the nuclear motion over a square grid.~\cite{SOREVIK200956,mclachlan2006geometric} Details over the models are provided in Section~2 and in the Supplementary information. One quantum feature that is not incorporated in the model is the nuclear permutation symmetry in H$_2^+$. We do not distinguish the behavior of the para- and ortho- Hydrogen varieties. We comment some possible effects in the Conclusions. As a precautionary principle, and for comparison, in this work we also study the rotational and vibrational dynamics in the excited state of HD$^+$, where nuclear spin symmetry plays no role. In addition, HD$^+$ presents a permanent nuclear dipole that competes against the transient dipole, putting in perspective the pervasiveness of anti-alignment dynamics. Section~3 shows the results of our calculations in both molecules using both models, for molecules initially aligned or misaligned with the field. Finally, Section~4 is the Conclusions.

\section{Hydrogen molecular ions under a strong field}

In this section, we intend to design two models for the dynamics of the different hydrogen molecular ions, under a strong field, that gives an accurate description of the time-evolution of both nuclear and electronic wave functions. In the first model, we combine the Ehrenfest approach for nuclear motion with the well established prolate spheroidal treatment for an electron in a two-center molecule.~\cite{baik1996multiphoton,kamta2004high,kamta2005three} Let us consider two Hydrogen isotopes $(a)$ and $(b)$ with masses $m_a$ and $m_b$ and an electron. The nuclear and electronic motion follows the time-dependent Schr\"odinger equation (TDSE), in atomic units,
\begin{align} \label{eq:tdse}
  i \frac{\partial }{\partial t} \Psi(\textbf{R}, \textbf{r}, t) =&  H  \Psi(\textbf{R}, \textbf{r}, t) \nonumber \\
  =& - \frac{1}{2 \mu_{ab}} \nabla_{\textbf{R}}^2 \, \Psi(\textbf{R}, \textbf{r}, t) - \frac{1}{2} \nabla_{\textbf{r}}^2 \, \Psi(\textbf{R}, \textbf{r}, t) \nonumber \\
  &+ \left[V(\textbf{R}, \textbf{r}) + E \, z - \frac{1}{2} \alpha E  R \cos \theta \right] \Psi(\textbf{R}, \textbf{r}, t) \ ,
\end{align}
where $H$ is the Hamiltonian, $\textbf{r} = x \, \hat x + y \, \hat y + z \, \hat z$ the spatial coordinate of the electron relative to the geometric center of the nuclei, $\textbf{R}$ the internuclear vector (pointing from the nucleus ($b$) to the nucleus ($a$)), $E$ the electric field magnitude, $\theta$ the angle between $R$ and $\hat z$, the reduced mass $\mu_{ab} = (m_a m_b) / (m_a + m_b)$, the asymmetry parameter\cite{esry1999adiabatic} $\alpha = (m_b - m_a) / (m_a + m_b)$, and $V(\textbf{R}, \textbf{r})$ the potential which is given by
\begin{equation*}
  V(\textbf{R}, \textbf{r}) = - \frac{1}{r_a} - \frac{1}{r_b} + \frac{1}{R} \ ,
\end{equation*}
where $r_{\alpha}$ is the distance between the electron and the nuclei $\alpha$ $(\alpha = a,b)$. For the sake of simplicity, we obviate the contribution of the motion of the center of mass to the dynamics, which represents a phase factor that does not change the observables, and can be taken away by a unitary transformation of the wave function. In the same lines, we have chosen without loss of generality $\hat z$ as the orientation of the electric field. Note that the asymmetry leads to a permanent dipole moment linearly increasing with the internuclear distance that tends to align the lighter nucleus towards the field. In the following, we will assume $(a)$ as the lighter nucleus. On the other hand, if $\alpha = 0$, there is no permanent dipole.  Asymptotically, as the ground and first excited state become degenerate, there is a charge transfer state due to the intrinsic ambiguity of which atom does the electron reside that leads to a linearly increasing polarizability with opposite signs for the ground and excited state, responsible for the different alignment dynamics.

\subsection{Ehrenfest model}

In the Ehrenfest model, we solve the nuclear motion following an ensemble of semi-classical trajectories starting from a Wigner distribution. For each trajectory, we solve the Schr\"odinger equation for the electronic motion, namely
\begin{align*} \label{eq:He}
  i \frac{\partial }{\partial t} \psi(\textbf{r}) &=  H_e(R, \theta, E) \psi(\textbf{r}) \nonumber \\
  &= - \frac{1}{2} \nabla_{\textbf{r}}^2 \, \psi(\textbf{r}) + \left[V(\textbf{R}, \textbf{r}) + E \, z - \frac{1}{2} \alpha E  R \cos \theta \right] \psi(\textbf{r}) \ ,
\end{align*}
using the prolate spheroidal treatment for an electron in a two-center molecule. This allows us to write Hamiltonian matrix elements over an orthonormal basis. At the same time, we incorporate the nuclear motion by the Hellmann-Feynman force approximation, namely
\begin{equation} \label{eq:HF1}
  \frac{d^2 R}{d t^2} = R \left(\frac{d \theta}{d t}\right)^2  - \frac{1}{\mu_{ab}} \Big< \psi \Big|\frac{\partial \check H_e}{\partial R} \Big| \psi \Big>
\end{equation}
and
\begin{equation} \label{eq:HF2}
  \frac{d^2 \theta}{d t^2} = - \frac{2}{R} \left(\frac{d R}{d t}\right) \left(\frac{d \theta}{d t}\right) - \frac{1}{\mu_{ab} R^2} \Big< \psi \Big|\frac{\partial \check H_e}{\partial \theta} \Big| \psi \Big> \ .
\end{equation}
Note that fictional forces appear because we are using two generalized coordinates, the angle $\theta$ between the molecule and the field, and the internuclear distance $R$. It is important to understand that, although one can consider the dynamics as three dimensional, the momentum conservation along the perpendicular axis to the plane of the molecule and the field, forces the nuclei to move within this plane (as long as the state of the electron is symmetric with respect such plane). Hence, we need only two generalized coordinates to describe the motion of both nuclei around its center of mass.

Finally, we use the fourth order Runge-Kutta method to propagate the electronic probability amplitudes along the nuclear motion. We typically perform 400 simulations to represent the behavior of an ensemble of molecules, all starting with different initial conditions from a random sampling of the Wigner distribution that we will detail further in the manuscript.

\subsection{Fully quantum 3+2D model}

In the second model, we solve the (3+2)TDSE, with three coordinates for the electron motion and two others for the nuclear motion, namely, the internuclear distance $R$ and the angle between the molecular axis and the field, $\theta$. The (3+2)TDSE reads
\begin{multline} \label{eq:(3+2)0}
  i \frac{\partial }{\partial t} \Psi(\textbf{R}, \textbf{r}) =  \left[- \frac{1}{2 \mu_{ab} R} \frac{\partial }{\partial R} R \frac{\partial }{\partial R} - \frac{1}{2 \mu_{ab} R^2} \frac{\partial^2}{\partial \theta^2} \right. \\ \left. - \frac{1}{2} \nabla^2_{\textbf{r}} + V(\textbf{R}, \textbf{r})  + E z - \frac{1}{2} \alpha E R \cos \theta\right]  \, \Psi(\textbf{R}, \textbf{r}) \ .
\end{multline}
Expanding the wave function in terms of the electronic states $\psi_n$ in the following way
\begin{equation*}
  \Psi(\textbf{R}, \textbf{r}) = \sum_{n} \sqrt{R} \, \phi_{n}(R, \theta) \, \psi_{n} (\textbf{r}) \ ,
\end{equation*}
Eq. \eqref{eq:(3+2)0} becomes
\begin{multline} \label{eq:nuclear}
  i \frac{\partial }{\partial t} \phi_n(R, \theta) = \left[- \frac{1}{2 \mu_{ab}} \frac{\partial^2 }{\partial R^2} - \frac{1}{8 \mu_{ab} R^2} - \frac{1}{2 \mu_{ab} R^2} \frac{\partial^2}{\partial \theta^2}\right] \phi_n(R, \theta) \\ + \sum_{n'} \big<\psi_n\big| H_e \big| \psi_{n'} \big>  \, \phi_{n'}(R, \theta) \ ,
\end{multline}
where $\phi_n$ is the nuclear wave function in the electronic state $\psi_n$. We define $\psi_n$ from the eigenfunctions of the electronic Hamiltonian $H_e$ without the field, $\varphi_n$,
\begin{equation*}    
    H_e(R, \theta, 0) \, \varphi_n(R;\textbf{r}) = H_{e, n} \, \varphi_n(R;\textbf{r}) \ ,
\end{equation*}
in the following way
\begin{equation*}    
    \psi_n(\textbf{r}) = \varphi_n\left(R_0; \textbf{r} \frac{R}{R_0}\right) \ ,
\end{equation*}
with $R_0 = 2$ a$_0$. The electronic states $\psi_n$ are therefore scaled eigenfunctions of the electronic Hamiltonian. Consequently, the matrix that represents this operator is not diagonal. The advantage is that we only need to diagonalize it once in the simulation, and the matrix of the Hamiltonian turns out to be very sparse, increasing the computational speed. Nevertheless, in the following analysis, when we refer to the ground or excited state, we will mean the actual eigenstates of the electronic Hamiltonian $\varphi_n$. More details are available on the supplementary information.

Finally, we solve Eq. \eqref{eq:nuclear} propagating all nuclear wave functions $\phi_n$ over a grid of 256 points between $0.1$ a$_0$ and $50$ a$_0$ for the $R$ coordinate, and 128 points between $0$ and $2 \pi$ for the $\theta$ coordinate, using a Strang method with a three operator-splitting scheme\cite{SOREVIK200956,mclachlan2006geometric}.

\begin{figure*}
  \centering
  \includegraphics[scale=0.68]{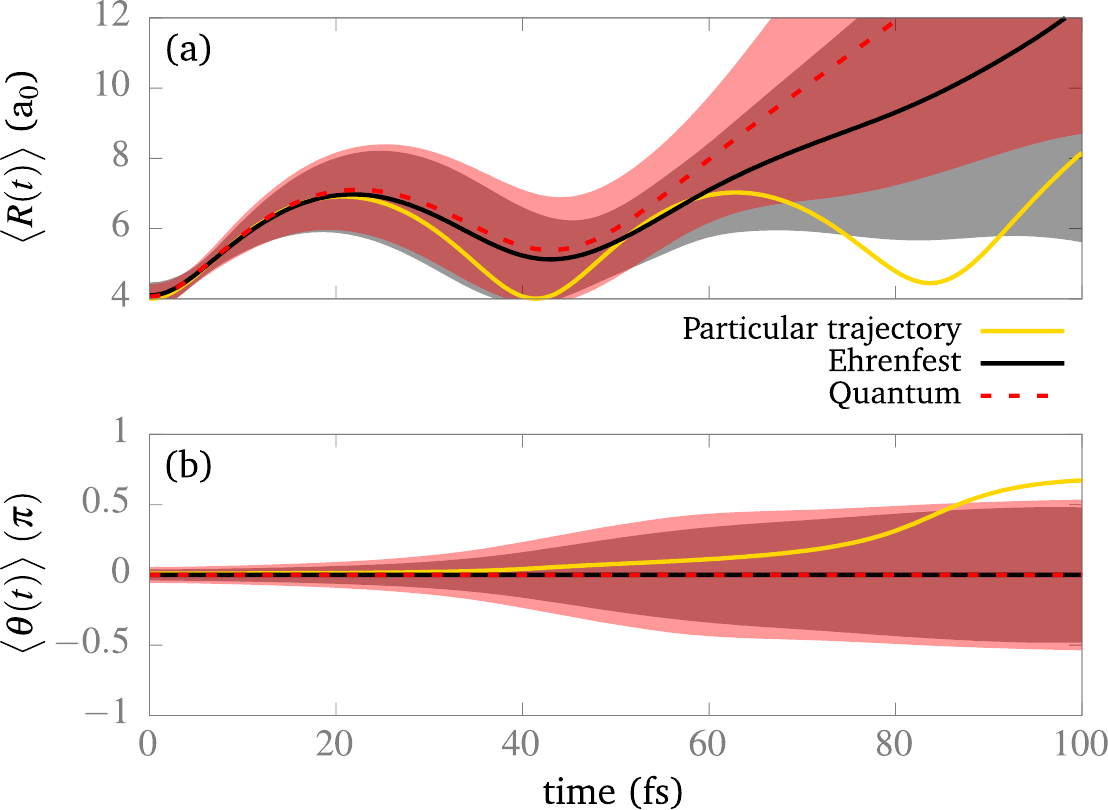}
  \hspace{1 cm}
  \includegraphics[scale=0.68]{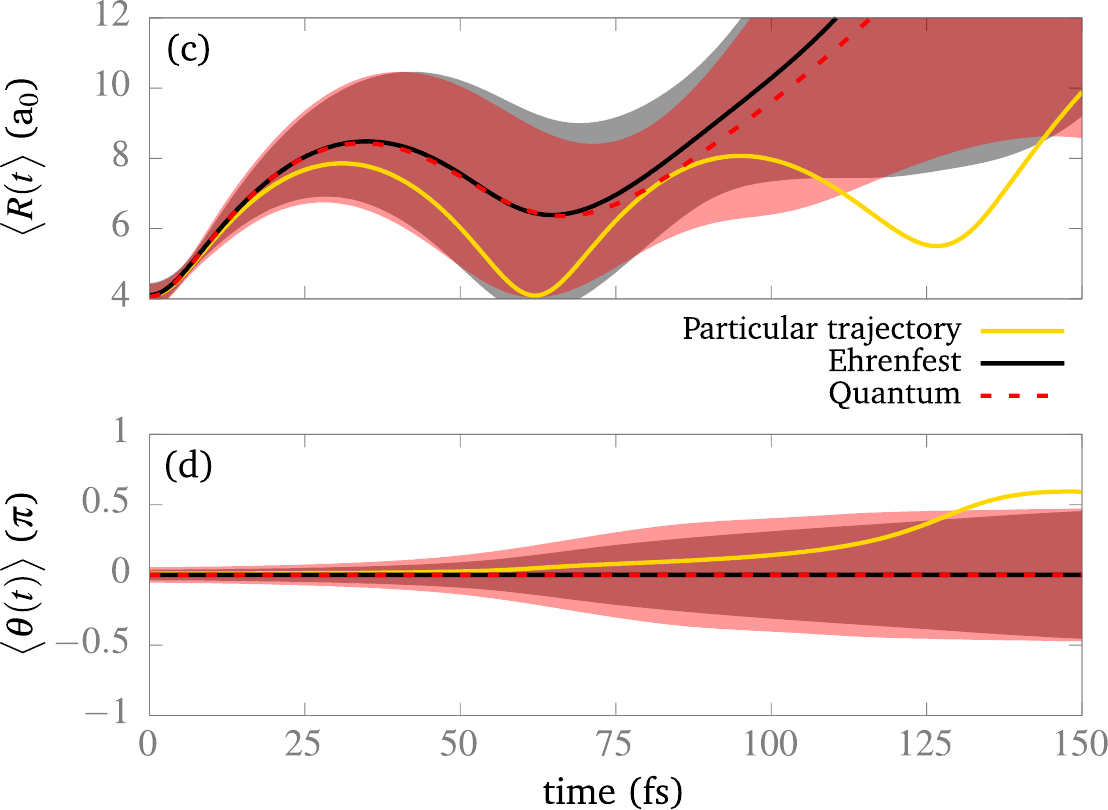}
  \caption{Time evolution of the internuclear distance $\big<R(t)\big>$ for (a) the H$_2^+$ molecule and (c) the HD$^+$ molecule, and the expected values of the angle between the field and the molecular axis $\big< \theta(t)\big>$ for (b) the H$_2^+$ molecule and (d) the HD$^+$ molecule. In all cases, we plot the corresponding standard deviation in filled curves with the same color code. For both approaches we assume as the initial condition a two-dimensional Gaussian wave packet  in the first excited electronic state (initially aligned with the field). In the Ehrenfest approach, we follow the time evolution of an ensemble of $400$ molecules according to the corresponding Wigner distribution. The field is constant with $E = 0.02$ a.u. The golden line show a particular trajectory from the ensemble. \label{fig:anti-align}}
\end{figure*}

\section{Results}

We first study the dynamics in the first excited state, in the presence of a strong field, for both H$_2^+$ and HD$^+$. In homonuclear diatomics, it is well known that the transient dipole increases with the internuclear distance (due to a charge transfer resonance), leading to bond hardening\cite{ChangJPB15,ZuoPRAR95,NiikuraPRL04}. On the other hand, using soft-core Coulomb potentials within a semi-classical Ehrenfest model, it was shown that the molecular axis aligns perpendicular to the field polarization. In spite of all the approximations of the model, the physical effect behind the anti-alignment could be traced to a quadratic increase in the polarizability and the change in its sign with respect to that of the ground state\cite{chang2019control}. As a first application of the methods developed in this work, we will test to which degree these approximations affect the results by gradually relaxing or removing the approximations.

In Fig. \ref{fig:anti-align}, we show the average and standard deviations in $R$ and $\theta$ both for the fully quantum results and the Ehrenfest model, as the H$_2^+$ in the excited electronic state evolves under a field of amplitude $E = 0.02$ a.u., polarized in the $\hat z$ direction. For the initial state, we assume a nuclear wave packet aligned with the field, $\langle \theta(0) \rangle = 0$, with an angular dispersion of $\sigma_\theta \approx 0.17$ (in radians), consistent with experimental conditions~\cite{mizuse2015quantum}. Regarding the internuclear distance, we choose $\langle R(0) \rangle = 4$ a$_0$ and $\sigma_R \approx 0.36$ a$_0$, which approximately are the conditions expected if an initial H$_2$ molecule in the ground state is ionized and then excited  with a pump pulse of $\lambda = 800$ nm, as used in Ref.\cite{chang2019control}. In the Ehrenfest model, we first obtain the Wigner distribution related to the wave packet and then generate an ensemble of trajectories. Care must be taken when computing the standard deviation and other statistic features of an angular distribution. In this work, we have used the standard methods of directional statistics to define such quantities correctly. \cite{fisher1995statistical}

As expected, the Ehrenfest calculations underestimate the degree of vibrational dephasing at longer times, as the wave packet always moves in a single (albeit highly anharmonic) FIP. However, one observes qualitatively the same behavior, particularly at earlier times. Indeed, in Fig. \ref{fig:anti-align}(a), we observe from both models a partial vibration in the FIP followed by dissociation. At the same time, in Fig. \ref{fig:anti-align}(b), we see that although the molecular alignment remains on average, the standard deviation of the orientation grows along with the dynamics, thus showing anti-alignment. Within the Ehrenfest model, a particular trajectory starting with the molecule almost aligned [$\theta = \pi/100$ rad. in golden line in Fig. \ref{fig:anti-align}(a)] clearly reveals how $\theta =0$ is an unstable critical point. As time evolves, the molecular axis slowly turns perpendicular to the field. As the coupling with the field weakens, the vibration in the FIP softens and the molecule finally dissociates.

Now we turn our attention to the HD$^+$ molecule, where the mass difference not only alters the timescale of the dynamics but creates a permanent dipole that tends to align the molecule with the field. In Fig. \ref{fig:anti-align} (c) and (d), we show the results of the dynamics under the same initial conditions and electric field as in the previous case (aligned nuclear wave packet in the excited state with $\sigma_\theta \approx0.17$ rad., $\langle R(0) \rangle = 4$ a$_0$, $\sigma_R \approx 0.36$ a$_0$). Surprisingly, despite the permanent dipole, we observe anti-alignment dynamics again. However, we observe the consequences of the permanent dipole in the period of the partial vibration in the FIP. Both Eq. \eqref{eq:HF1} and \eqref{eq:HF2} suggest that the period in the HD$^+$ case should be $\sqrt{4/3}$ times the period in the H$_2^+$ case. However, the permanent dipole diminishes the bond hardening as it leads to a potential linearly decreasing with the internuclear distance, generating slower vibrations. 
As a consequence, the vibrational dephasing is smaller and the Ehrenfest dynamics follows more closely the fully quantum case.

\begin{figure}
  \centering
  \includegraphics[scale=0.68]{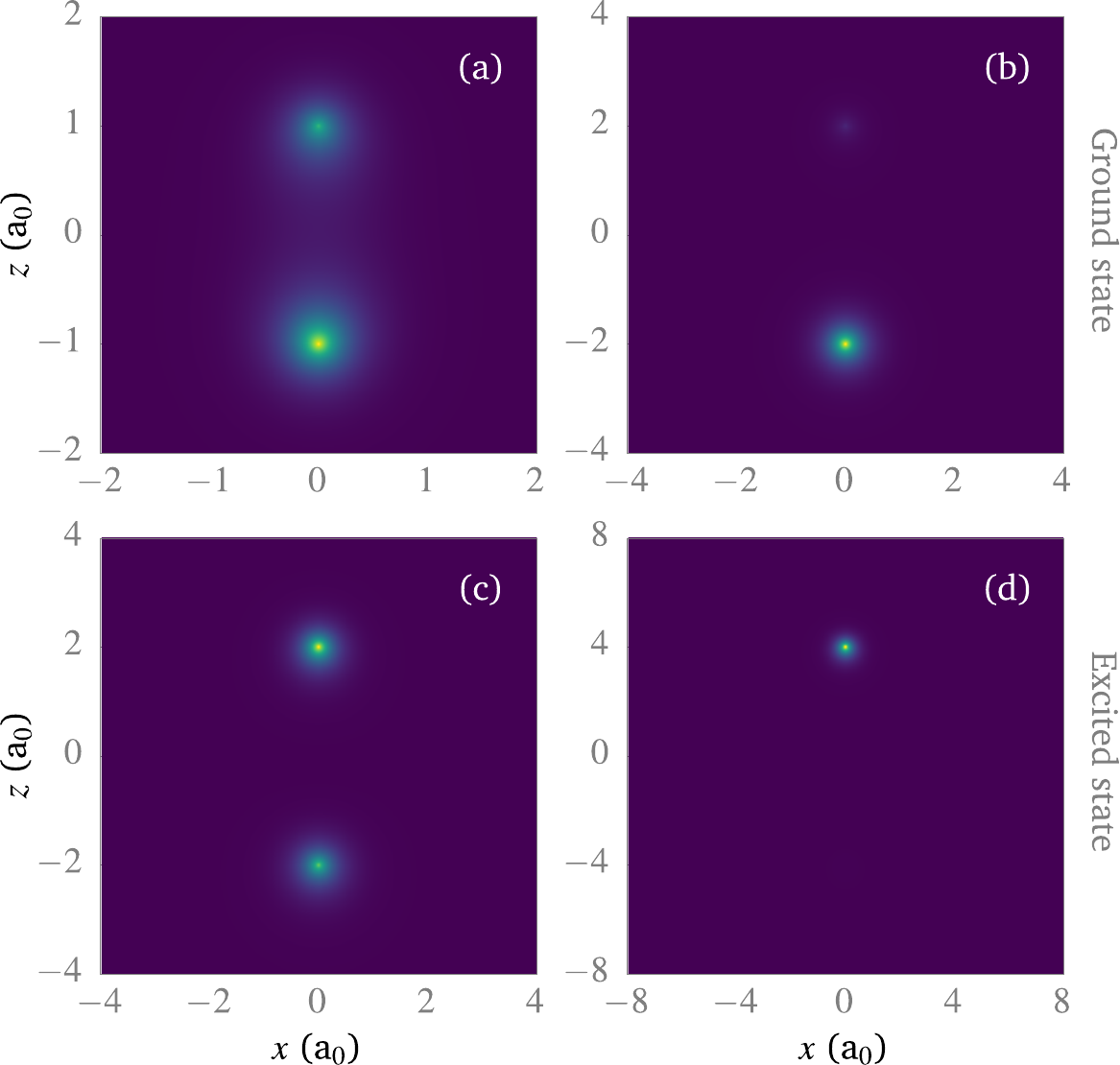}
  \caption{Electronic density of the H$_2^+$ and HD$^+$ molecules aligned with an electric field, for the ground and excited states, at different fixed internuclear distances $R$, namely, (a) at the ground state with $R=2$ a$_0$, (b) at the ground state with $R=4$ a$_0$, (c) at the first excited state with $R=4$ a$_0$, and (d) at the first excited state with $R=8$ a$_0$. \label{fig:electronic_states}}
\end{figure}

To understand why in spite of a permanent dipole, the HD$^+$ molecule rotates against the field, we show in Fig. \ref{fig:electronic_states} the ground and excited electronic states of the molecule, for a fixed internuclear distance. As noticed, the electron becomes localized around the nuclei against the field when the molecule is in the ground electronic state, especially for relatively large internuclear distances. Such electronic density is responsible for the emergence of a dipole that points towards the field and the consequent alignment dynamics in the ground electronic state. On the contrary, we observe a dipole against the field when the molecule is in the excited electronic state, which surpasses the permanent dipole and triggers the anti-alignment dynamics.

\begin{figure}
  \centering
  \includegraphics[scale=0.68]{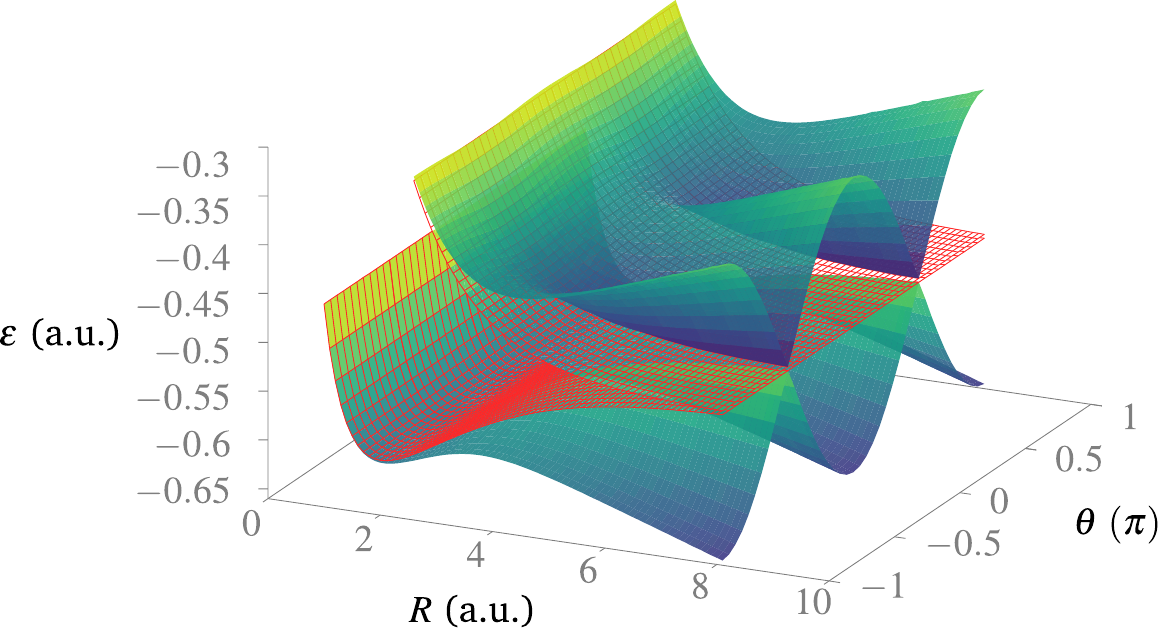}
  \caption{Field induced potential energy surface of the H$_2^+$ molecule, under a constant electric field $E = 0.02$ a.u., in the ground and excited electronic state. At $\theta = \pm \pi/2$, the potentials are the true molecular potentials (without any field, shown in red for reference purposes), allowing the transition between the ground and first excited electronic states. \label{fig:conica}}
\end{figure} 

\begin{figure*}
    \centering
        \includegraphics[width=0.45\textwidth]{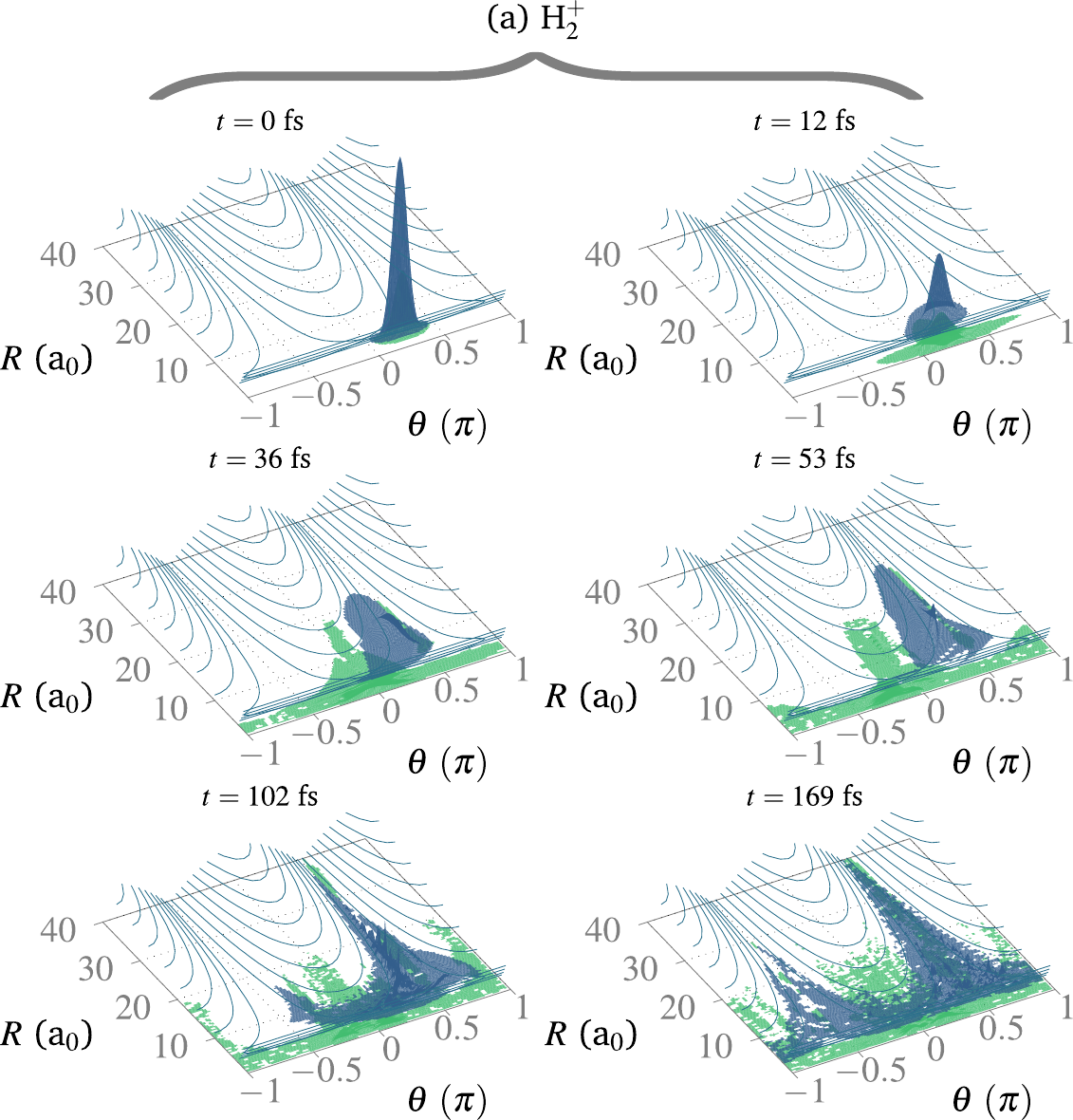}
        \qquad
        \includegraphics[width=0.45\textwidth]{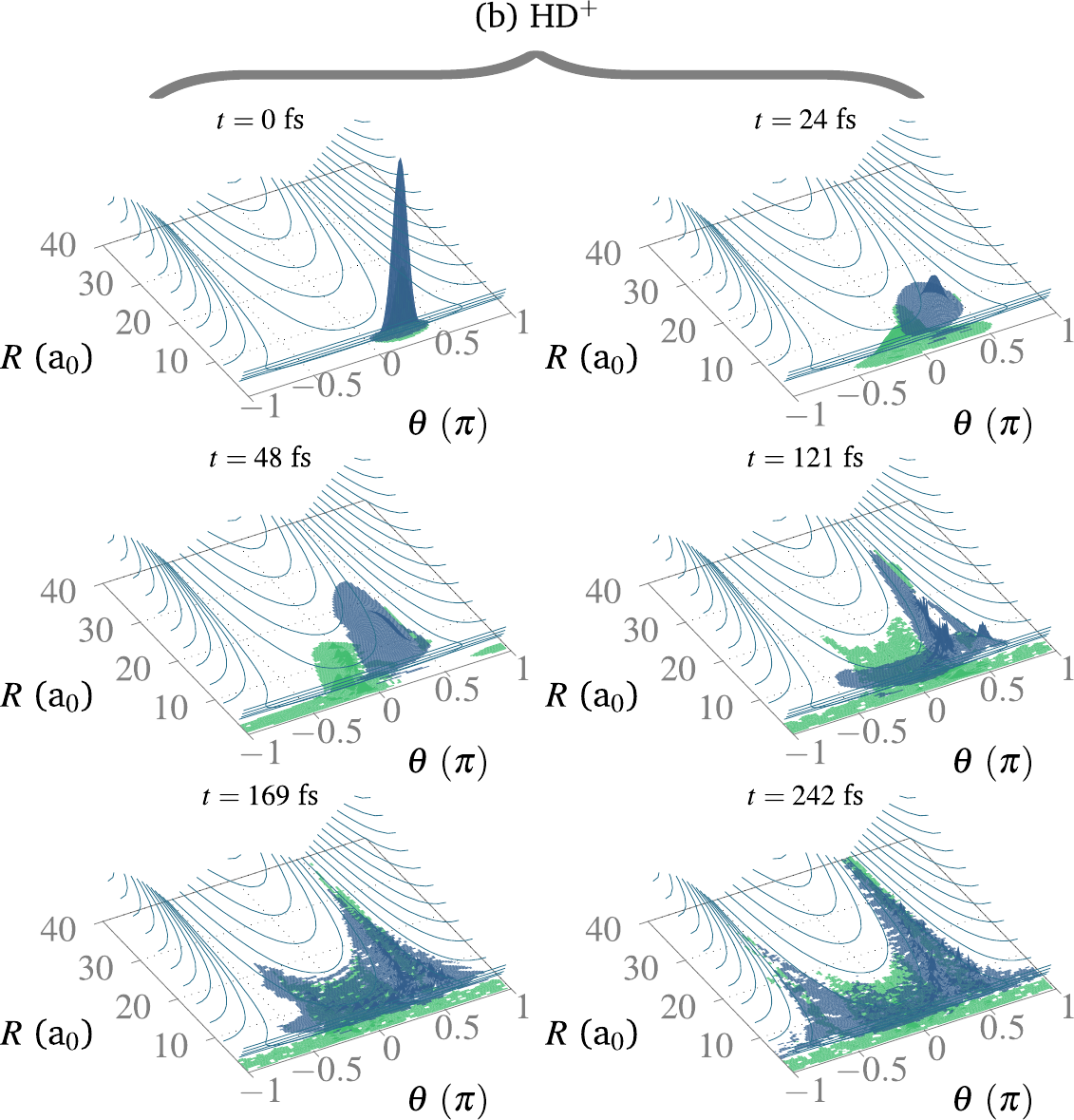}
    \caption{Time evolution of a Gaussian nuclear wavepacket initially
      misaligned with the electric field ($\langle \theta \rangle = \pi/4$) in a
      superposition of the ground (green) and the first excited
      electronic state (blue), for the H$_2^+$ molecule (a) and the
      HD$^+$ molecule (b), according to the fully quantum
      model. The first excited state LIP is in blue lines.}\label{fig:timev}
\end{figure*}

Finally, we study the dynamics when the molecule is initially misaligned with the field, using the fully quantum model. Here we want to focus not only on the vibrational and rotational degrees of freedom, but also on the electronic populations. Since anti-alignment drives the molecular axis perpendicular to the field, at large internuclear distances both FIPs approach, as shown in Fig. \ref{fig:conica}, such that the nuclear motion may allow the electronic transition in a way similar to how non-adiabatic couplings induce transitions between Born-Oppenheimer electronic states. When one uses an electromagnetic field to control the process, the potentials cross at a given internuclear distance where $V_e(R)-\hbar\omega = V_g(R)$ forming a true LICI.
We want to understand the dynamics around these points and the contribution of dissociation in the excited state versus in the ground electronic state, due to bond softening. To do so, we plot in Fig. \ref{fig:timev} the time evolution of the probability of finding (a) the H$_2^+$ molecule and (b) the HD$^+$ molecule oriented with an angle $\theta$ to the field with internuclear distance $R$, in the ground electronic state (green) and the first excited electronic state (blue), starting from a superposition of both the ground and first excited electronic states. For the initial nuclear wave packet we choose the same features as previously $\langle R \rangle = 4$ a$_0$, $\sigma_R \approx 0.36$ a$_0$,  $\sigma_\theta \approx 0.17$ rad., but with $\langle \theta \rangle = \pi/4$. In the supplementary information, we provide the movies of the nuclear dynamics, which are more illustrative than the figures. As observed, the projection of the nuclear wave function on the excited state (the excited packet) first elongates due to the weaker bond hardening at the given angle, and then anti-aligns, allowing for the subsequent dissociation in the excited state. On the other hand, because bond softening in the ground potential is also weaker at $\langle \theta \rangle = \pi/4$ than for the aligned molecule, the projection of the nuclear wave function on the ground state (the ground packet) shrinks and then elongates as the ground  packet aligns with the field, where bond softening is stronger. 

As the excited packet crosses $\theta = \pi/2$ there is some population transfer 
to the ground state (the net balance is approximately a $5$\% population gain in the ground state), where it arrives at high energy and very quickly disperses for all values of $\theta$. The large gain of kinetic energy allows all orientations to be finally present, but still showing mainly alignment. Part of the ground packet dissociates, not at $\theta = 0$ (due to bond softening), but at $\theta = \pi/2$, as some components are above the dissociation limit. On the other hand, parts of the excited packet do not dissociate directly but vibrate and disperse, moving to $\theta = -\pi/2$ where again one can observe some (smaller) probability of dissociation in the excited state. At larger times all the population in the excited state will dissociate (at $\theta = \pi/2$ or $-\pi/2$) and part of the ground state will also dissociate.

\section{Conclusions}
We developed two models to assess the anti-alignment dynamics of the H$_2^+$ molecule (or its isotopes) in the excited state, where we incorporate the nuclear motion quantum mechanically or semi-classically through the Ehrenfest approach. In both cases we took into account the true Coulomb interactions between particles and all electronic degrees of freedom. The two models agree, especially at shorter times, and predict anti-alignment dynamics for the H$_2^+$ in the excited state, with respect to the field polarization, thus agreeing with our previous results from simplified models. Surprisingly, our results show the same prediction for the HD$^+$ molecule despite its permanent dipole. 
To explain these results, we computed the true adiabatic potentials in the presence of the field, so-called FIPs, and observed that although the bond hardening takes place in the excited state, it does not provide rotational stability to the aligned state. Indeed, the electronic density shows the emergence of an induced dipole against the field, which is ultimately responsible for the anti-alignment. 
We suspect that similar effects may affect the duration of strong-field effects on the excited states, not necessarily dissociative, of more complex molecules.

Both molecules dissociate at angles perpendicular to the field polarization. The dissociation is complete in  approximately two rotational periods and occurs both in the excited and in the ground potential, as the population is exchanged mediated by the nuclear motion, when the FIPs become closer at large internuclear distances in anti-aligned molecules.  Therefore, bond softening in the ground electronic state does not play any significant role in the dissociation.

Finally, the quantum mechanical dispersion of the molecular axis in the initial state is sufficient to cause full dissociation. 
Consequently, the stabilization of these molecules in the excited state through the bond-hardening produced by a constant electric field is highly unlikely.

One of the few effects that our quantum model has not incorporated is the nuclear spin, which has some important consequences on the rotational states in H$_2^+$, namely, only ortho-Hydrogen can exist in the excited electronic state with $J=0$. Therefore, we expect different dynamical behavior for para-Hydrogen and ortho-Hydrogen in an experiment. However, as our study in HD$^+$ shows, the physics is driven by the polarizability. Since all even-$J$ to odd-$J$ rotational transitions between the excited and ground electronic states are symmetry-allowed, the polarizabilities remain basically the same and in the presence of high intense fields, as in our work, the strong rotational coupling should allow the population of many rotational levels. Therefore, we do not expect that the discrete nature of the levels plays such an important role, although this should be a matter of further study.
Finally, additional dynamical effects caused by ultrashort and strong laser pulses and rotational excitation, will be explored in future studies.

\section{Acknowledgments}

This work was supported by CONICYT-PCHA/Doctorado Nacional/2016-21161403 (S.C.), Fondo Nacional de Investigaciones Cient\'ificas y Tecnol\'ogicas (FONDECYT, Chile) under Grants \#1190662 (J.R.), \#1190703 (J.A.V.), CEDENNA through the ``Financiamiento Basal para
Centros Cient\'ificos y Tecnol\'ogicos de Excelencia-FB0807" (S.C., J.R., J.A.V), and MINECO CTQ2015-65033-P (I.R.S.).

\bibliographystyle{apsrev4-1}
\bibliography{main.bib}

\end{document}